\PassOptionsToPackage{svgnames}{xcolor}
\documentclass[%
12pt
]{iopart}

\usepackage{hhline}

\usepackage{tikz}
\usetikzlibrary{plotmarks}
\usetikzlibrary{decorations.markings}

\usepackage{xcolor}

\usepackage{graphicx,caption,subcaption}
\usetikzlibrary{decorations.pathmorphing}

\usepackage{dcolumn}
\usepackage{bm}

\usepackage{hyperref}

\usepackage{mwe}

\usepackage{caption}
\usepackage{subcaption}

\renewcommand{\thesection}{}
\renewcommand{\thesubsection}{\arabic{section}.\arabic{subsection}}
\makeatletter
\def\@seccntformat#1{\csname #1ignore\expandafter\endcsname\csname the#1\endcsname\quad}
\let\sectionignore\@gobbletwo
\let\latex@numberline\numberline
\def\numberline#1{\if\relax#1\relax\else\latex@numberline{#1}\fi}
\makeatother

\usepackage{blindtext}
\usepackage[T1]{fontenc}
\usepackage[utf8]{inputenc}
\usepackage[icelandic,english]{babel}

\usepackage{titlesec}
\titleformat{\section}
  {\normalfont\sffamily\large\bfseries\color{black}}
  {\thesection}{1em}{}

\hypersetup{colorlinks=true,urlcolor=blue}

\renewcommand{\thesubsection}{\Alph{subsection}}

\usepackage{etoolbox}
\makeatletter
\def\@mkboth#1#2{}
\newlength\appendixwidth
\preto\appendix{\addtocontents{toc}{\protect\patchl@section}}
\newcommand{\patchl@section}{%
  \settowidth{\appendixwidth}{\textbf{Appendix }}%
  \addtolength{\appendixwidth}{1.5em}%
  \patchcmd{\l@section}{1.5em}{\appendixwidth}{}{\ddt}%
}
\makeatother

\usepackage{iopams}

\expandafter\let\csname equation*\endcsname\relax

\expandafter\let\csname endequation*\endcsname\relax

\usepackage{amsmath}

\usepackage{dcolumn}
\usepackage{epsf}
\usepackage{float}

\usepackage{tabularx}

\usepackage{mathtools}

\DeclarePairedDelimiter\floor{\lfloor}{\rfloor}

\usepackage[final]{changes}

\usepackage{lipsum}
\definechangesauthor[color=blue]{.}
\colorlet{Changes@Color}{red}
\setremarkmarkup{(#2)}

\usepackage[export]{adjustbox}

\usepackage{datetime}

\usepackage[cal=boondoxo]{mathalfa}

\allowdisplaybreaks
\begin{document}

\title[]{Chebyshev Polynomial Expansion of Two-Dimensional Landau-Fermi liquid Parameters
%
}

\author{Joshuah T. Heath, Matthew P. Gochan, \& Kevin S. Bedell}

\address{Physics Department,  Boston  College,  Chestnut  Hill, Massachusetts  02467,  USA}
\ead{heathjo@bc.edu}
\vspace{10pt}
\begin{indented}
\item[]\today
\end{indented}

\begin{abstract}  
We study the intrinsic effects of dimensional reduction on the transport equation of a perfectly two-dimensional Landau-Fermi liquid. By employing the orthogonality condition on the 2D analog of the Fourier-Legendre expansion, we find that the equilibrium and non-equilibrium properties of the fermionic system differ from its three-dimensional counterpart, with the latter changing drastically. Specifically, the modified Landau-Silin kinetic equation is heavily dependent on the solution of a non-trivial contour integral specific to the 2D liquid. We find the solution to this integral and its generalizations, effectively reducing the problem of solving for the collective excitations of a collisonless two-dimensional Landau-Fermi liquid to solving for the roots of some high-degree polynomial. This analysis ultimately lays the mathematical foundation for the exploration of atypical behavior in the non-equilibrium properties of two-dimensional fermionic liquids in the context of the Landau quasiparticle paradigm.
\end{abstract}

\vspace{2pc}

\submitto{\JPA}

\maketitle

\tableofcontents

%


\section{Introduction}

It is widely known that dimensional reduction often has a dramatic effect on the mathematical description of a physical system. 
A reduction to two dimensions has been well-studied in conformal field theory, where the 2D conformal Killing equations reduce to the Cauchy-Riemann equations \cite{Senechal,Schottenloher}. Perhaps more dramatically, the Einstein-Hilbert action in two dimensions is a topological invariant, leading to the breakdown of conventional methods in general relativity and the non-existence of a well-defined Newtonian limit\cite{Jackiw,Witten,Romero}. 
%
 Although the particularities of these two examples are difficult to confirm experimentally, the conclusions drawn are built into the mathematics of two-dimensional Lie algebra and classical field theory, respectively, and give us strong examples where dimensional reduction results in an {\it a priori} sense of physical intuition. 

Non-trivial behavior inherent to two-dimensional systems is similarly seen in many-body and solid state physics, where various 2D materials such as graphene \cite{Novoselov,Kotov,Pablo} and GaAs-AlGaAs heterojunctions under intense magnetic fields \cite{Tsui1,Tsui2} exhibit highly unconventional collective phenomena. Nevertheless, despite its relevance to similar materials, a rigorous exploration of the consequences of dimensional reduction inherent to the 2D neutral Landau-Fermi liquid is severely lacking. In its original formulation, Fermi liquid theory provides a phenomenological approach to interacting Fermi systems by assuming an isomorphism between the eigenstates of the non-interacting and interacting systems \cite{Landau1,Landau2}. Although highly successful in describing the normal state of most metals in the absence of topological order, 
in one dimension an alternative formulation (the Tomonaga-Luttinger model) must instead be taken, which uses bosonization techniques to describe the universal low-frequency/long-wavelength behavior of the system \cite{Haldane_Luttinger_liquid}. In two dimensions, however, the fate of the Landau-Fermi liquid phenomenology has yet to be universally agreed upon, despite extensions of perturbative techniques hinting at a stable ground state \cite{Serene, Coffey} and non-trivial behavior unseen in three dimensions \cite{Farinas}. The study of 2D Fermi liquids has been primarily limited to microscopic analyses of systems in the strong hydrodynamic regime \cite{DasSarma2,DasSarma3}. Applications of higher-dimensional bosonization appear to suggest an "orthogonality catastrophe" in 2D Fermi liquids in the presence of an impurity \cite{Anderson1, Anderson2, Anderson3, Anderson4}, while effective mass calculations in the random-phase approximations appear to suggest a possible Mott transition or Wigner crystallization in 2D electron systems \cite{DasSarma1}. Where a phenomenological approach is taken, three-dimensional quantities are automatically assumed to apply to the 2D case \cite{Grimmer}. 

In this work, we show that the effects of dimensional reduction have a highly non-trivial, intrinsic effect on 
the non-equilibrium properties of the two-dimensional Fermi liquid. Our argument hinges upon the widespread assumption that the distortion of the Fermi surface is independent of the magnitude of the quasiparticle momentum \cite{Landau1, Landau2,Baym}. In the 3D case, one usually expands fundamental quantities in terms of normalized spherical harmonics, subsequently leading us to the quantification of collective modes in terms of different oscillations of the Fermi surface characterized by the degree $\ell$ of the corresponding Legendre polynomial $P_\ell(\cos(\theta))$. The main results of 3D Landau-Fermi liquid theory follows from the orthogonality condition on Legendre polynomials\cite{Baym}:
\begin{align}
\int_{-1}^1 P_{\ell}(x)P_{\ell'}(x)dx=\frac{\delta_{\ell \ell'}}{2\ell+1}
\end{align}
 In the 2D case we consider in this paper, we implement an analogous study by invoking the Chebyshev polynomial $T_\ell(\cos\phi)$ \cite{Chebyshev, Rivlin, Mason}. Much as the Legendre polynomials form an orthogonal basis in 3D, the Chebyshev polynomials form an orthogonal basis in two dimensions. However, the orthogonality condition for these polynomials takes the form
\begin{align}
\int_{-1}^1 \frac{T_\ell(x)T_{\ell'}(x)}{\sqrt{1-x^2}}dx =\frac{\pi}{2} \delta_{\ell \ell'}\left(
1+\delta_{\ell 0}
\right)
\end{align}
The inclusion of the additional multiplicative term in the left-hand side of the above has drastic consequences on the non-equilibrium behavior of a two-dimensional Fermi liquid in the collisionless limit, mostly due to the contribution of the weighting factor unseen in the 3D Legendre orthogonality condition. Ultimately, we find that the dispersion for zero sound has a closed form if we consider all Landau parameters $F_{\ell>0}=0$. The inclusion of higher-order Landau parameters up to degree $\ell$ reduces the problem of solving for the collective mode dispersion to solving for the roots of a polynomial of degree $2\ell+1$. As such, the mathematical formalism introduced in this article lays the groundwork for the study of atypical behavior in the dispersion relation of 2D zero sound \cite{Gochan}.


%
%

%
%

\section{Equilibrium properties of the 2D Landau-Fermi liquid}

\subsection{Pomeranchuk instability condition of the 2D Landau-Fermi liquid}

Due to the fact that the Landau parameter is only dependent on the relative angle between $p$ and $p'$, the effect of the Legendre polynomials and their orthogonalitiy arises prominently in the calculation of the Pomeranchuk instability condition \cite{Pomeranchuk} and effective mass \cite{Landau1} of a three-dimensional Fermi liquid. In order to understand the role of the Chebyshev polynomial on the equilibrium properties of the 2D system (and how it differs from the higher-dimensional case), we briefly review in this section their explicit derivation. 

Let us first make sure we understand the Pomeranchuk instability condition of the 3D Fermi liquid. The free energy functional is given by
\begin{align}
\delta F=F-F_0=\frac{1}{V}\sum_{p\sigma} (\epsilon_{p\sigma}-\mu) \delta n_{p\sigma} +\frac{1}{2V^2}\sum_{\substack{pp'\\ \sigma \sigma'}}f_{\sigma \sigma'}(p,\,p')\delta n_{p\sigma}\delta n_{p'\sigma'}
\end{align}
where $F_0$ is the free energy of the non-interacting system and we take the usual truncation to quadratic order in the change in the distribution function $\delta n_{p\sigma}=n_{p\sigma}-n_{p\sigma}^0$. Linearizing the dispersion in the vicinity of the Fermi surface and performing a Taylor expansion of the Heaviside step function, we find the following simplification:
\begin{align}
\delta F
&=\frac{v_F p_F^2}{(2\pi \hbar)^3} \sum_{\ell m} |u_{\ell m}|^2 \left(
1+\frac{F_\ell^s}{2\ell+1}
\right)
\end{align}
Because $\delta F>0$ for the Fermi liquid phase to remain stable, we arrive at the Pomeranchuk instability condition for a 3D Fermi liquid \cite{Pomeranchuk, Baym}:
\begin{align}
1+\frac{F_\ell^s}{2\ell+1}>0
\end{align}
For the $\ell=0$ channel, this reduces to $F_0^s>-1$. Any Landau parameter $F_0^s$ smaller than negative one will therefore lead to $\delta F<0$ and the system will not support a Fermi liquid-like ground state. 

When considering the 2D Landau-Fermi liquid, the term of $\delta F$ linear in $\delta n_{p\sigma}$ is initially changed slightly due to dimensional reduction:

\begin{align}
v_F\sum_{p\sigma}(p-p_F)\delta n_{p\sigma} 
\approx\frac{v_Fp_F}{(2\pi\hbar)^2}\int d\Omega \, \delta p_F^2 
\end{align}
However, unlike the 3D case, we must now perform a mode expansion in terms of the Chebyshev polynomial if we are to fully consider the effects of phase space reduction:

\begin{align}
\int d\Omega \,\delta p_F^2&=\sum_{\substack{\ell_1 \ell_2}} u_{\ell_1} u_{\ell_2}\int d\Omega\, T_{\ell_1}(\phi) T_{\ell_2}(\phi)
\end{align}
By expanding the Chebyshev polynomial in terms of exponential functions, one can show that the integral above is given by the following:
\begin{align}
\int_0^{2\pi} d\Omega \, T_{\ell_1}(\phi)T_{\ell_2}(\phi)&=\begin{cases}
0,\quad \ell_1\not =\ell_2 \label{eqn13}\\
\pi,\quad \ell_1 =\ell_2\not=0\\
2\pi,\quad \ell_1 =\ell_2 =0
\end{cases}=\pi \delta_{\ell_1 \ell_2}(1+\delta_{\ell_1 0})
\end{align}
Hence, the linear term of the Landau free energy is given by
\begin{align}
v_F \sum_{p\sigma}(p-p_F)\delta n_{p\sigma}&\approx \frac{\pi v_F p_F}{(2\pi \hbar)^2}\sum_{\ell } |\nu_{\ell }|^2(1+\delta_{\ell 0})
\end{align}
The quadratic term must be expanded in a similar fashion: 
\begin{align}
\frac{1}{2V^2}\sum_{\substack{pp' \\ \sigma \sigma '}}f_{\sigma \sigma'}(p,\,p')\delta n_{p\sigma}\delta n_{p'\sigma'}
&=\frac{2p_F^2}{(2\pi\hbar)^4} \sum_{\ell\ell_1\ell_2}\nu_{\ell_1}\nu_{\ell_2}f_\ell^s \int_0^{2\pi} d\phi \int_0^{2\pi} d\phi' \cos(\ell_1\phi) \cos(\ell_2 \phi') \cos(\ell |\phi-\phi'|)\notag\\
&=\frac{2p_F^2}{(2\pi\hbar)^4} \sum_{\ell\ell_1\ell_2}\nu_{\ell_1}\nu_{\ell_2}f_\ell^s\int_0^{2\pi} d\phi\,T_{\ell_1} (\phi) T_\ell (\phi) \int_0^{2\pi} d\phi' T_{\ell_2}(\phi')T_{\ell}(\phi')\notag\\
&=\frac{2p_F^2\pi^2}{(2\pi \hbar)^4} \sum_\ell |u_\ell|^2 f_\ell^s (1+\delta_{\ell 0})^2
\end{align}
By once again invoking the dimensionless form of the Landau parameter, we find that the free energy is then given by

\begin{align}
\delta F
&=\frac{\pi v_F p_F}{(2\pi \hbar)^2} \sum_{\ell m} |u_{\ell}|^2 \bigg(
1+\delta_{\ell 0}+\frac{F_\ell^s}{2} (1+\delta_{\ell 0})^2
\bigg)
\end{align}
Because $\delta F>0$ for stability of the Fermi liquid ansatz to make sense, we find the 2D variant of the Pomeranchuk instability condition:
\begin{align}
1+\frac{F_\ell^s}{2}(1+\delta_{\ell 0})\ge 0\label{eq14}
\end{align} 
When $\ell=0$ in the above, we have the same stability condition as in the 3D system; i.e., that $F_0^s>-1$. However, Eqn. \eqref{eq14} tells us that the condition for stability is $F_\ell^s>-2$ for all $\ell>0$. This is a direct result of expanding the Fermi surface distortion in terms of two-dimensional Chebyshev polynomials, as opposed to expanding in terms of three-dimensional Legendre polynomials.

\subsection{Effective mass of the 2D Landau-Fermi liquid}

One can check the instability condition derived in the previous section by considering the dependence of the 2D effective mass on the Landau parameters. By enforcing the Landau-Fermi liquid to be Galilean invariant and defining the quasiparticle current to be ${\bf j}_k={\bf k}/m$, we find the relationship between the quasiparticle mass $m^*$ and the bare mass $m$ to be 
\begin{align}
\frac{m^*}{m}=1+\frac{1}{V v_F^*}\sum_{{\bf k}'\sigma'}f_{\sigma \sigma'}({\bf k},\,{\bf k}')\delta(k-k_F) \cos(\theta)
\end{align}
By expanding the Landau parameter in terms of Legendre polynomials, we find a factor of $1/3$ in the above \cite{Landau1, Baym}.
In the two dimensional Fermi liquid, the argument from Galilean invariance remains. However, the effects of dimensional reduction and Chebyshev polynomial orthogonality yields a factor of $1/2$ as opposed to a $1/3$ in the 3D system:

\begin{align}
\frac{1}{V v_F^*}\sum_{k'\sigma'}f_{\sigma \sigma'}({\bf k},\,{\bf k}')\delta(k-k_F)\cos\phi&=\frac{2}{(2\pi\hbar)^2}\frac{1}{v_F^*}\int d\Omega \int dk'\,k'\,f_{\sigma \sigma'}({\bf k},\,{\bf k}')\cos(\phi)\delta (k-k_F)\notag\\
&=F_\ell^s\int d\phi\, T_\ell( \phi)\cos(\phi)\notag\\
&=\frac{F_1^s}{2}
\end{align}
Hence, we see that for both the Pomeranchuk instability condition and the effective mass, the 2D results differ from their 3D analogs by numerical constants. The underlying physics of these equilibrium properties therefore remain practically the same under dimensional reduction, up to some numerical constants.



\section{The Landau kinetic equation in two dimensions}

\subsection{The role of the Chebyshev Polynomial in 2D linearized transport equation}

In the collisionless regime (i.e., $\omega\tau>>1$), the linearized kinetic equation may be written in the form 
\begin{align}
\left(
\frac{\partial}{\partial t}+{\bf v}_p\cdot \nabla 
\right)\delta n_{\bf p}({\bf r},\,t)-\left(
\frac{\partial n_{{\bf p}}^0}{\partial \epsilon_{\bf p}}
\right){\bf v}_p\cdot \nabla \delta \epsilon_{\bf p}({\bf r},\,t)=0
\end{align}
where $\delta n_{\bf p}$ is the change in the quasiparticle distribution function $n_{\bf p}$, $v_{\bf p}$ is the quasiparticle velocity, and $\delta \epsilon_{\bf p}({\bf r},\,t)$ is the effective field given by
\begin{align}
\delta \epsilon_{\bf p}({\bf r},\,t)=U({\bf r},\,t)+\sum_{{\bf p}'}f_{pp'}\delta n_{{\bf p}'}({\bf r},\,t)
\end{align}

When we consider the linearized transport equation of the 2D equation, the underlying structure before the expansion in terms of partial waves is identical to the 3D case \cite{Baym}:

\begin{align}
\nu_p+\frac{ {\bf q}\cdot {\bf v}_p}{\omega-{\bf q}\cdot {\bf v}_p}\sum_{p'\sigma'}f_{pp'}^{\sigma \sigma'}\frac{\partial n_{p'}^0}{\partial \epsilon_{p'}^0}\nu_{p'}=\frac{{\bf q}\cdot {\bf v}_p}{\omega-{\bf q}\cdot {\bf v}_p}U
\end{align}

We now perform the following expansions:
\begin{align}
f_{pp'}^{s,\,a}=\frac{1}{\sqrt{2\pi}} f_0^{s,\,a}+\frac{1}{\sqrt{\pi}}\sum_{\ell=1} T_\ell (\cos\theta) f_\ell^{s,\,a}
\end{align}
\begin{align}
\nu_p=\frac{1}{\sqrt{2\pi}}\nu_0+\frac{1}{\sqrt{\pi}}\sum_{\ell=1}T_\ell(\cos\theta)\nu_\ell
\end{align}
Note that we have invoked the dimensional reduction of the available phase space by replacing the inherently three-dimensional Legendre polynomial $P_\ell(x)$ with the two-dimensional Chebyshev polynomial $T_\ell(x)$. 

Direct substitution yields
\begin{align}
\sum_\ell T_{\ell}(x)\nu_\ell+\frac{x}{s-x}\sum_{p'\sigma'}\sum_{\ell''}T_{\ell''}(x'')f_{\ell''}^s \left(\frac{\partial n_{p'}^0}{\partial \epsilon_{p'}}\right)\sum_{\ell'}T_{\ell'}(x')\nu_{\ell'}=\frac{x}{s-x}U
\end{align}
where we have taken the usual substitution $x=\cos\theta$ and $s=\omega/qv_F$. The second expression in the above may be simplified as follows:

\begin{align}
\sum_{p'}\sum_{\ell''}T_{\ell''}(x'')f_{\ell''}^s \left(\frac{\partial n_{p'}^0}{\partial \epsilon_{p'}}\right)\sum_{\ell'}T_{\ell'}(x')\nu_{\ell'}&=\int_0^{2\pi}d\theta'\sum_{\ell''}T_{\ell''}(x'')f_{\ell''}^s \left(
-\frac{2m^*}{(2\pi\hbar)^2}f_{\ell''}^s
\right)\sum_{\ell'}T_{\ell'}(x')\nu_{\ell'}\notag\\
&=-\frac{2m^*}{(2\pi\hbar)^2}\sum_{\ell'',\,\ell'}\int_0^{2\pi}d\theta' f_{\ell''}^sT_{\ell''}(x'')\  T_{\ell'}(x')
\end{align}
After calculating the integrals in the above, the 2D Landau kinetic equation can be cast into the form
\begin{align}
\sum_\ell T_\ell(x) \nu_\ell -\frac{1}{2}\frac{x}{s-x}\sum_{\ell'}F_{\ell'}^sT_{\ell'}(x)\nu_{\ell'}(1+\delta_{\ell' 0})=\frac{x}{s-x}U
\end{align}
We can now exploit the well-known orthogonality condition central to the study of Chebyshev polynomials:
\begin{align}
\int_{-1}^1 \frac{T_\ell(x)T_{\ell'}(x)}{\sqrt{1-x^2}}dx =\begin{cases}
\frac{\pi}{2}\delta_{\ell \ell'},\,\quad &\ell\not =0,\quad \ell'\not =0\notag\\
\pi,\quad &\ell=\ell'=0
\end{cases}
\end{align}
Which yields 
\begin{align}
\sum_\ell \int_{-1}^1 \frac{T_\ell(x)T_{\ell}(x)}{\sqrt{1-x^2}}\nu_\ell dx
-\frac{1}{2}\int_{-1}^1 \frac{x}{s-x}\sum_{\ell'}(1+\delta_{\ell'0})F_{\ell''}^s \frac{T_{\ell'}(x)T_{\ell}(x)}{\sqrt{1-x^2}}dx=\int_{-1}^1 \frac{x}{s-x}U\frac{T_\ell(x)}{\sqrt{1-x^2}}dx\label{7}
\end{align}
It is the evaluation of the integrals in the above that will be the central focus of this paper.


\subsection{$\ell=0$ Response Function}

Recalling Eq. $\eqref{7}$, we will derive the most easily-attainable response function; namely, where only the $\ell=0$ channel contributes significantly. Making this assumption, we immediately obtain
\begin{align}
\nu_0 \int_{-1}^1 \frac{dx}{\sqrt{1-x^2}}-F_0^s\nu_0 \int_{-1}^1 \frac{x}{s-x}\frac{1}{\sqrt{1-x^2}}=U\int_{-1}^1 \frac{x}{s-x}\frac{1}{\sqrt{1-x^2}}dx
\end{align}
The first integral trivially evaluates to $\pi$, yielding the following for the response function:
\begin{align}
\frac{\nu_0}{U}&=\frac{\int_{-1}^1 \frac{x}{s-x}\frac{dx}{\sqrt{1-x^2}}}{\pi -F_0^s \int_{-1}^1 \frac{x}{s-x}\frac{dx}{\sqrt{1-x^2}}}\notag\\
&=\left(\left(
\frac{1}{\pi}\int_{-1}^1 \frac{x}{s-x}\frac{dx}{\sqrt{1-x^2}}\right)^{-1}-F_0^s\right)^{-1}
\end{align}
where we have made the ansatz that the integral is never zero. We are therefore left with calcuating the following integral:
\begin{align}
\mathcal{I}_1\equiv 
\frac{1}{\pi}\int_{-1}^1 \frac{x}{s-x}\frac{dx}{\sqrt{1-x^2}}\label{10}
\end{align}
For future reference, we will call this the {\bf Landau-Chebyshev integral of the first order}.

To find the poles of the response function, we will have no choice but to solve Eqn. \eqref{10} for the general case of some complex form of $s$. Although we may use the well-known Sokhotski-Plemelj theorem for $s$ defined on the real line, we wish to derive the general form of the integral for some complex $s$, and so we instead turn to straight contour integration. At first, we are tempted to take the standard "dog bone" contour to evaluate this integral (see Figure 1), which is defined by taking a branch cut for $x\in [-1,1]$. Although this works for $|s|>1$, we face immediate difficulties if $s\in [-1,1]$, as then the pole lies in the branch cut. However, what is less apparent is that we also face similar issues for some general complex $s$ whenever $|s|<1$. To see this, note that the residue of Eqn. \eqref{10} for some general complex pole $s$ will lie in the interval $[-1,1]$ whenever $|s|<1$. By definition, the legitimacy of the residue theorem for a given analytic function depends upon the ability to take a Laurent expansion of said function. Because the square root is not fully analytic in a punctured disc about the branch cut, the Laurent expansion fails about the branch cut. Therefore, if the residue falls in the branch cut, we have taken the wrong contour for our problem, and we must be careful how we approach the integral for general $s$.

\subsection{Contour approach to the First Order Landau-Chebyshev integral}
%
%
%

\begin{figure}
\begin{center}
\begin{tikzpicture}[
    scale=3,
    line cap=round,
    dec/.style args={#1#2}{
        decoration={markings, mark=at position #1 with {#2}},
        postaction={decorate}
    }
]
\path [gray,thin] (-1.2,0) edge[->] (1.2,0)  (0,-1.2) edge[->] (0,1.2);
\draw [blue, thick, dec={0.29}{\arrow{>}}, dec={0.82}{\arrow{>}}]
    (-.08,1)coordinate(31) arc (95:445.8:1cm) coordinate(13); 

\draw [black, dashed, dec={0.29}{--}] (0,0) circle (0.5 cm); 


\draw [red, thick, xshift=.5cm, dec={0.29}{\arrow{<}}]
    (-160:1mm)coordinate(21) arc (-160:160:1mm) coordinate(11); 
\draw [red, thick, xshift=-.5cm, dec={0.29}{\arrow{<}}]
    (20:1mm)coordinate(12)   arc (20:340:1mm) coordinate(22);    
\draw [red,thick ]  (11)--(.08,.034) 
			   (-.08,.034)--(12)   
			   (-.08,-.034)--(22) 
			   (-.08,-.034)--(21)
			   (-0.08, 0.034)--(-0.08,1)
			   (0.08, 0.034)--(0.08,1);

\path (.29*360:1.15cm) node {}
(.31*360:.4cm) node {}
    (-5mm,2mm) node {}
    (5mm,2mm) node {}
        (5mm,6mm) node {$s_1$}
                (2mm,3.1mm) node {$s_2$};
\end{tikzpicture}
\end{center}
\caption{The "dog bone" contour that is commonly used to solve integrals of a similar form to Eqn. \eqref{10}. The blue contour is taken to infinity, while the red contour is infinitely tightened. When the pole $s\equiv s_1$ is outside the unit circle (dashed in black), the residue theorem may be applied. However, when the pole $s\equiv s_2 \not\in [-1,1]$ is inside the unit circle, the corresponding residue of Eqn. \eqref{10} lies along the branch cut.}

\begin{center}
\begin{tikzpicture}[
    scale=3,
    line cap=round,
    dec/.style args={#1#2}{
        decoration={markings, mark=at position #1 with {#2}},
        postaction={decorate}
    }
]
\path [gray,thin] (-1.2,0) edge[->] (1.2,0)  (0,-1.2) edge[->] (0,1.2);

\draw [orange, thick, dec={0.39}{\arrow{>}},dec={0.89}{\arrow{>}}] (0,0) circle (0.5 cm); 



\path (.29*360:1.15cm) node {}
(.31*360:.4cm) node {}
    (-5mm,2mm) node {}
    (5mm,2mm) node {}
            (.1mm,.1mm) node {$z_1$}
        (5mm,6mm) node {$z_2$}
                (2mm,3.1mm) node {$z_3$};
\end{tikzpicture}
\end{center}
\caption{The simple contour that results from a change of variables for Eqn. \eqref{10}. By representing the integrand as a rational function, the poles can be easily read off for any general value of $s$ without having to worry about branch cuts on the real axis.}
\end{figure}

\indent 

Because we wish to derive a formula for the first order Landau-Chebyshev integral that is applicable to all complex $s$ (and in preparation for dealing with Landau-Chebyshev integrals of higher order), we make a change of variables to circumvent this issue. Let $x=\cos(\theta)$. Then Eqn. \eqref{10} becomes
\begin{align}
\frac{1}{\pi}\int_{-1}^1 \frac{x}{s-x}\frac{dx}{\sqrt{1-x^2}}&=\frac{1}{\pi}\int_{-\pi}^0 \frac{\cos\theta}{s-\cos\theta}\frac{d(\cos\theta)}{\sqrt{1-\cos^2\theta}}\notag\\
&=\pm\frac{1}{\pi}\int_0^\pi \frac{\cos\theta}{\cos\theta-s}d\theta\notag\\
&=\pm \frac{1}{\pi}\int_\pi^{2\pi} \frac{\cos\theta}{\cos\theta-s}d\theta
\end{align}
This tells us that
\begin{align}
\frac{1}{\pi}\int_{-1}^1 \frac{x}{s-x}\frac{dx}{\sqrt{1-x^2}}=-\frac{1}{2\pi}\int_0^{2\pi}\frac{\cos\theta}{\cos\theta-s}d\theta
\end{align}
where we have taken the negative sign to match the trivial result for $|s|>1$. We now take a second change of variables, letting $\cos\theta=\frac{z+z^{-1}}{2}$:
\begin{align}
\frac{1}{\pi}\int_{-1}^1 \frac{x}{s-x}\frac{dx}{\sqrt{1-x^2}}&=-\frac{1}{\pi}\frac{1}{2}\int_0^{2\pi}\frac{\cos\theta}{\cos\theta-s}d\theta\notag\\
&=-\frac{1}{2\pi}\oint_{|z|=1}\frac{\frac{z+z^{-1}}{2}}{\left(\frac{z+z^{-1}}{2}\right)-s}\frac{dz}{iz}\notag\\
&=-\frac{1}{2\pi i}\oint_{|z|=1}\frac{z^2+1}{z^3-2sz^2+z}dz\notag\\
&=-\frac{1}{2\pi i}\oint_{|z|=1}\frac{(z-i)(z+i)}{z(z-s-\sqrt{s^2-1})(z-s+\sqrt{s^2-1})}
\end{align}

%
%
%
%
\noindent 
\noindent We denote the poles of the above as $z_1=0$, $z_2=s+\sqrt{s^2-1}$, and $z_3=s-\sqrt{s^2-1}$. Because the contour is taken with $|z|=1$ (see Fig. 2), the pole at $z_1=0$ always contribute to the final expression. The other two are somewhat nontrivial. We can split these two up into their real and imaginary components quite easily if $|s|>1$:
\begin{align}
z_2=s'+\sqrt{|s|^2-1}+is'',\quad z_3=s'-\sqrt{|s|^2-1}+is''
\end{align}
Note that, throughout this paper, we will take $s'\equiv \Re (s)\in \mathbb{R}$ and $s''\equiv \Im (s)\in \mathbb{R}$. Hence,
\begin{align}
|z_2|&=\sqrt{\left(s'+\sqrt{|s|^2-1}\right)^2+s''^2}\notag\\
&=\sqrt{2\left(s^2+s'\sqrt{|s|^2-1}\right)-1}
\end{align}
\begin{align}
|z_3|&=\sqrt{\left(s'-\sqrt{|s|^2-1}\right)^2+s''^2}\notag\\
&=\sqrt{2\left(s^2-s'\sqrt{|s|^2-1}\right)-1}
\end{align}
For $|z_2|$, assuming $s'>0$, we can clearly see that $|z_2|>1$, and therefore $z_2$ is not included in the contour. For the third term, we will now prove that $|z_3|$ is always contained in the contour for some $s'>0$ and $|s|>1$. Namely, we want to show that
\begin{align}
|s^2-s'\sqrt{|s|^2-1}|<1
\end{align}
For any small $s'$, we can clearly see that the above is satisfied, as $|s|^2>1$. Logically, the maximum value of $s'=s$; i.e., $s''=0$. This tells us that the above expression becomes
\begin{align}
s(s-\sqrt{|s|^2-1})<1
\end{align}
This is what we want to prove. Because $s$ is completely real, $s>1$. Hence, it should be clear that
\begin{align}
s-\sqrt{|s|^2-1}<1
\end{align}
The above equality is clearly true for $|s|>1$. Thus, $z_3$ is contained within the contour and $z_2$ is not if we assume $|s|>1$.
We can the calculate the integral using the residue theorem:
\begin{align}
&\phantom{=}\pm\frac{1}{2\pi i}\oint_{|z|=1}\frac{(z-i)(z+i)}{z(z-s-\sqrt{|s|^2-1})(z-s+\sqrt{|s|^2-1})}\notag\\
&=\pm \bigg\{
\frac{1}{\left(
-s-\sqrt{|s|^2-1}
\right)\left(
-s+\sqrt{|s|^2-1}
\right)}+
\frac{(s-\sqrt{|s|^2-1})^2+1}{\left(
s-\sqrt{|s|^2-1}
\right)\left(
-2\sqrt{|s|^2-1}
\right)}
\bigg\}\notag\\
&=\pm \left\{
1-\frac{s}{\sqrt{|s|^2-1}}
\right\}
\end{align}
It is important to note that, for the case when $s'=0$ with $|s|>1$, then
\begin{align}
|z_2|=|z_3|=\sqrt{2|s|^2-1}
\end{align}
For this scenario, because $s^2>1$, neither of the poles is included in the contour, and the integral reduces to $\pm 1$. 

We will now consider the case where $|s|<1$. We then find that
\begin{align}
z_2=s'+i\left(
s''+\sqrt{1-|s|^2}
\right),\quad z_3=s'+i\left(
s''-\sqrt{1-|s|^2}
\right)
\end{align}
We can easily see that
\begin{align}
|z_2|&=\sqrt{s'^2+(s''+\sqrt{1-|s|^2})^2}\notag\\
&=\sqrt{1+2s''\sqrt{1-|s|^2}}
\end{align}

\begin{align}
|z_3|&=\sqrt{s'^2+(s''-\sqrt{1-|s|^2})^2}\notag\\
&=\sqrt{1-2s''\sqrt{1-|s|^2}}
\end{align}

For both of the above, $|s|<1$ and, therefore, $|s''|<1$. Assuming $s''$ is positive, then $|z_2|>1$ and $|z_3|<1$, much as before:

\begin{align}
&\phantom{=}\pm\frac{1}{2\pi i}\oint_{|z|=1}\frac{(z-i)(z+i)}{z(z-s-\sqrt{|s|^2-1})(z-s+\sqrt{|s|^2-1})}dz
=\pm \left\{1-\frac{s}{\sqrt{|s|^2-1}}\right\}
\end{align}
This is true for either zero or finite $s'$. However, if $s''=0$, then we have the more interesting case that $|z_2|^2=|z_3|^2=1$. Because this is on the contour, we can change the contour such that there is a small semicircle of radius $\epsilon$ around the pole:
\begin{align}
&\phantom{=}\pm\frac{1}{2\pi i}\oint_{|z|=1}\frac{(z-i)(z+i)}{z(z-s-\sqrt{|s|^2-1})(z-s+\sqrt{|s|^2-1})}dz\notag\\
&=\frac{1}{2\pi i}\int_0^{\pi} \lim_{\epsilon\rightarrow 0}\left\{\frac{(\epsilon e^{i\theta}+1)^2+1}{
(\epsilon e^{i\theta}+1)(
\epsilon e^{i\theta}+1 -s -\sqrt{|s|^2-1}
)(
\epsilon e^{i\theta}+1-s+\sqrt{|s|^2-1}
)
}i\epsilon e^{i\theta}\right\}d\theta \notag\\
&=\frac{1}{2\pi }\int_0^{\pi}\lim_{\epsilon\rightarrow 0}\left\{\frac{2 e^{i \theta } s \epsilon }{-2 e^{i \theta
   } (s-1) \epsilon -2 s+e^{2 i \theta } \epsilon
   ^2+2}-\frac{1}{1+e^{i \theta } \epsilon }+1\right\}d\theta\notag\\
   &=C
\end{align}
where $C$ is a constant.

The above analysis tells us the following. If we assume $|s|>1$, then
\begin{align}
&\phantom{=}\pm\frac{1}{2\pi i}\oint_{|z|=1}\frac{(z-i)(z+i)}{z(z-s-\sqrt{|s|^2-1})(z-s+\sqrt{|s|^2-1})}
=\begin{cases}
- 1+\frac{s}{\sqrt{|s|^2-1}},\quad &s'\not=0,\quad s''\not =0\notag\\
- 1+\frac{s}{\sqrt{|s|^2-1}},\quad &s'\not=0,\quad s'' =0\notag\\
- 1,\quad &s'=0,\quad s''\not=0\notag\\
\end{cases}
\end{align}

If we instead assume $|s|<1$, then 
\begin{align}
&\phantom{=}\pm\frac{1}{2\pi i}\oint_{|z|=1}\frac{(z-i)(z+i)}{z(z-s-\sqrt{|s|^2-1})(z-s+\sqrt{|s|^2-1})}
=\begin{cases}
-1+\frac{s}{\sqrt{|s|^2-1}},\quad &s'\not=0,\quad s''\not =0\notag\\
-1,\quad &s'\not=0,\quad s''=0\notag\\
-1+\frac{s}{\sqrt{|s|^2-1}},\quad &s'=0,\quad s''\not=0\notag\\
\end{cases}
\end{align}

\noindent where, without loss of generality, we take $C=0$. Therefore, we can clearly see that the integral is a constant for completely imaginary $s$ if $|s|>1$ and is a constant for completely real $|s|$ if $s<1$. We can write this compactly by introducing the following function:
\begin{align}
\eth(s)\equiv 1-\left\{
\delta(s')\Theta(|s|-1) +\delta(s'')\Theta(1-|s|)
\right\}
\end{align}
where the Icelandic letter ''eth'' is used for this special function. Hence, the closed form yields the final solution as
\begin{align}
\frac{1}{\pi} \int_{-1}^1 \frac{x}{s-x}\frac{dx}{\sqrt{1-x^2}}=\pm\left(
1-\frac{s}{\sqrt{s^2-1}}\textrm{\dh}(s)
\right)
\end{align}
The "eth" function $\eth(s)$ is then seen to take into account the finite-valuedness of $s'$ ($s''$) for $|s|>1$ ($|s|<1$). 

Returning to the form of the response function, we can now write down a final expression\footnote{Note that our ansatz that the integral is always non-zero is clearly observed in this expression.}: 

\begin{align}
\frac{\nu_0}{U}
&=\left( \left(
\frac{1}{\pi}\int_{-1}^1 \frac{x}{s-x}\frac{dx}{\sqrt{1-x^2}}\right)^{-1}-F_0^s\right)^{-1}\notag\\
&=\left(
\pm\left(
1-\frac{s}{\sqrt{s^2-1}}\textrm{\dh}(s)
\right)^{-1}-F_0^s
\right)^{-1}
\end{align}
Excitations of the system may be read off as poles of the response function. This is explored in detail in \cite{Gochan}, where the form of the first order Landau-Chebyshev integral guarantees the absence of Landau damping in a stable 2D Landau-Fermi liquid up to order $F_0^s$. From that study, one can clearly show that the negative branch of the square root is physical, as this agrees with the Pomeranchuk instability condition. Moreover, it is also important to note that the dispersion of zero sound which results form the above yields a closed solution {\it for all values of $s$}. Such a clean solution is a direct result of the Chebyshev orthogonality condition. This is vastly different from the 3D case, where the form of the Legendre polynomial tells us that the limit of large or small $F_0^s$ must be taken 
%
%
to achieve an approximate solution for the dispersion of the collective mode.

\section{Generalization of the Landau-Chebyshev integral for generic $\ell$}
\subsection{Zero sound response function for general $\ell$} 

In the previous section, we have shown that the response function for a 2D Fermi liquid takes a closed form for all $s$ if we exclude all higher-order $\ell>0$ channels. We will now show that the inclusion of all higher $\ell$ contributions to the Landau parameter $F_\ell^s$ will yield a similar solution by generalizing the Landau-Chebyshev integral to general $n$th order. 

In the collisionless limit, recall that the Landau-Chebyshev integral is given by

\begin{align}
&\sum_\ell\nu_\ell  \int_{-1}^1 \frac{T_\ell(x)T_{\ell}(x)}{\sqrt{1-x^2}}dx
+\frac{1}{2}\int_{-1}^1 \nu_{\ell'}\frac{x}{x-s}\sum_{\ell'}(1+\delta_{\ell'0})F_{\ell'}^s \frac{T_{\ell'}(x)T_{\ell}(x)}{\sqrt{1-x^2}}dx\notag\\
=&-U\int_{-1}^1 \frac{x}{x-s}\frac{T_\ell(x)}{\sqrt{1-x^2}}dx
\end{align}
From the orthogonality of the Chebyshev polynomials, the above simplifies to 
\begin{align}
\frac{1}{2}(1+\delta_{\ell 0}) \nu_\ell +\sum_{\ell'}\textrm{\th}_{\ell \ell'}(s)(1+\delta_{\ell' 0}) F_{\ell'}^s\nu_{\ell'} =-2\textrm{\th}_{\ell 0}(s)U
\end{align}
where we have defined the "thorn" function $\textrm{\th}_{\ell \ell'}$ to be
\begin{align}
\textrm{\th}_{\ell \ell'}(s)=\textrm{\th}_{\ell'\ell }(s)\equiv \frac{1}{2\pi }\int_{-1}^1 \frac{T_{\ell'}(x)T_{\ell}(s)}{\sqrt{1-x^2}}\frac{x}{x-s}dx
\end{align}
In this way, $\textrm{\th}_{\ell \ell'}$ is the two-dimensional analog of the generalized Lindhard function $\Omega_{\ell \ell'}$ which takes the full consideration of phase space reduction into account. Note that, even if we ignore contributions from $F_\ell^s$ when $\ell>0$, higher-order modes in $\textrm{\th}_{\ell \ell'}$ still become relevant to zero sound if we consider higher-order $\ell$ terms in the distortion of the 2D Fermi surface. This expression can easily be solved from the above
\begin{align}
\nu_\ell =\frac{2}{1+\delta_{\ell 0}}\frac{\textrm{\th}_{\ell 0}}{\textrm{\th}_{00}}\nu_0\label{eq43}
\end{align}
In a similar fashion, higher-order evaluations of the thorn integral are clearly relevant to the dispersion of first sound, although this is given by a more complex expression and requires the inclusion of the collision integral to make physical sense.

Unlike the generalized Lindhard function, we can exploit the mathematics of Chebyshev polynomials to reduce the task of finding the collective excitations of a 2D Fermi liquid to the problem of finding the roots of some general, finite $n$th order polynomial. To see this, note that any Chebyshev integral can be written as the following summation \cite{Rivlin, Mason}:
\begin{align}
T_n(x)=\sum_{k=0}^{\floor{\frac{n}{2}}}{n\choose 2k}(x^2-1)^k x^{n-2k}
\end{align}
Therefore, a product of Chebyshev polynomials can be written in the form
\begin{align}
T_{\ell'}(x)T_{\ell}(x)&=\sum_{k=0}^{\floor{\frac{\ell'}{2}}} {\ell' \choose 2k} (x^2-1)^k x^{\ell'-2k}\sum_{j=0}^{\floor{\frac{\ell}{2}}}{\ell \choose 2j}(x^2-1)^j x^{\ell-2j}\notag\\
&=\sum_{k=0}^{\floor{\frac{\ell'}{2}}}\sum_{j=0}^{\floor{\frac{\ell}{2}}} \sum_{m=0}^{k+j} {\ell' \choose 2k} {\ell \choose 2j} {k+j\choose m}(-1)^m x^{\ell+\ell'-2m}
\end{align}
The function $\textrm{\th}_{\ell \ell'}$ can then be written in the form
\begin{align}
\textrm{\th}_{\ell \ell'}(s)&=\frac{1}{2\pi}\int_{-1}^1 \frac{T_{\ell'}(x)T_{\ell}(x)}{\sqrt{1-x^2}}\frac{x}{x-s}dx\notag\\
&=\frac{1}{2\pi}\sum_{k=0}^{\floor{\frac{\ell'}{2}}}\sum_{j=0}^{\floor{\frac{\ell}{2}}} \sum_{m=0}^{k+j} {\ell' \choose 2k} {\ell \choose 2j} {k+j\choose m}(-1)^{m+1}\mathcal{I}_{\ell+\ell'-2m+1}
\end{align}
where we have introduced $\mathcal{I}_n$ to represent the $n$th order Landau-Chebyshev integral:
\begin{align}
\mathcal{I}_n=\frac{1}{\pi} \int_{-1}^1 \frac{x^n}{s-x}\frac{dx}{\sqrt{1-x^2}}
\end{align}
One can then solve for the sound dispersion in the collisionless limit for any order in the Landau parameter $F_\ell$ as long the above integral has a solution.

\subsection{Contour approach to the $n$th order Landau-Chebyshev integral}
Our goal is to now solve integrals of the form $\mathcal{I}_n$ given above. Our integral simplifies as before:

\begin{align}
\mathcal{I}_n&=\frac{1}{\pi}\int_{-1}^1 \frac{x^n}{s-x}\frac{dx}{\sqrt{1-x^2}}\notag\\
&=\pm \frac{1}{2\pi}\int_0^{2\pi} \frac{\cos^n\theta}{\cos\theta-s}d\theta\notag\\
&=\pm\frac{1}{2^n \pi i}\oint_{|z|=1}\frac{(z^2+1)^n}{z^n(z^2-2sz+1)}dz\notag\\
&=\pm\frac{1}{2^n \pi i}\oint_{|z|=1}\frac{(z^2+1)^n}{z^n(z-s-\sqrt{s^2-1})(z-s+\sqrt{s^2-1})}dz
\end{align}
Our integral therefore has the same three poles: $z_1=0$, $z_2=s+\sqrt{s^2-1}$, and $z_3=s-\sqrt{s^2-1}$. The only difference is now $z_1$ is an nth order pole. 

Our goal is to write down a closed solution for the integral for some general $n$, and thus we solve the contribution to each pole exactly. We will evaluate the $z_1$ pole first. From the residue theorem, the contribution from this pole yields solutions to the equation

\begin{align}
\frac{1}{2^{n-1}(n-1)!}\lim_{z\rightarrow 0}\frac{d^{n-1}}{dz^{n-1}}\left(
\frac{(z^2+1)^n}{(z-s-\sqrt{s^2-1})(z-s+\sqrt{s^2-1})}
\right)
\end{align}

This derivative can be simplified via a generalized product rule:
\begin{align}
\lim_{z\rightarrow 0}\frac{d^{n-1}}{dz^{n-1}}\left(
\frac{(z^2+1)^n}{z^2-2sz+1}
\right)
&=\sum_{k=0}^{n-1}{n-1 \choose k} \frac{d^{n-k-1}}{dz^{n-k-1}}(z^2+1)^n \frac{d^k}{dz^k}(z^2-2sz+1)^{-1}
\end{align}

The latter of the above derivatives can be solved by a double binomial expansion:
\begin{align}
\frac{d^k}{dz^k}\left( \frac{1}{z^2-2sz+1}\right)&=\frac{d^k}{dz^k}\sum_{j=0}^\infty (z^2-2sz)^j(-1)^j\notag\\
&=\frac{d^k}{dz^k}\sum_{j=0}^\infty z^j \sum_{\ell=0}^j {j \choose \ell}z^\ell (-2s)^{j-\ell}(-1)^j\notag\\
&=\sum_{j=0}^\infty \sum_{\ell=0}^j {j \choose \ell}(2s)^{j-\ell} (-1)^\ell \frac{(j+\ell)!}{(j+\ell-k)!}z^{j+\ell-k}
\end{align}
We will now perform the former derivative by a similar method:
\begin{align}
\frac{d^{n-k-1}}{dz^{n-k-1}}(z^2+1)^n=&\frac{d^{n-k-1}}{dz^{n-k-1}}\sum_{m=0}^{n}{n\choose m} z^{2m}\notag\\
=&\sum_{m=0}^n {n\choose m} \frac{(2m)!}{(2m+1-n+k)!}z^{2m+1-n+k}
\end{align}

Putting it all together, we have the following:

\begin{align}
&\lim_{z\rightarrow 0}\frac{d^{n-1}}{dz^{n-1}}\left(
\frac{(z^2+1)^n}{z^2-2sz+1}
\right)\notag\\
=&\lim_{z\rightarrow 0}\sum_{k=0}^{n-1}{n-1 \choose k} \frac{d^{n-k-1}}{dz^{n-k-1}}(z^2+1)^n \frac{d^k}{dz^k}(z^2-2sz+1)^{-1}\notag\\
=&\lim_{z\rightarrow 0}\sum_{k=0}^{n-1}{n-1\choose k} \left\{
\sum_{m=0}^n {n\choose m}\frac{(2m)!}{(2m+1-n+k)!}z^{2m+1-n+k}
\right\}\notag\\
&\phantom{
\sum_{k=0}^{n-1}{n-1\choose k}\,\,\,\,
}\times\left\{
\sum_{j=0}^\infty \sum_{\ell=0}^j (2s)^{j-\ell} (-1)^\ell {j \choose \ell}\frac{(j+\ell)!}{(j+\ell-k)!}z^{j+\ell-k}
\right\}\notag\\
=&\lim_{z\rightarrow 0}\sum_{k=0}^{n-1} \sum_{m=0}^n \sum_{j=0}^\infty \sum_{\ell=0}^j {n-1 \choose k} {n\choose m} {j\choose \ell} \frac{(j+\ell)!}{(j+\ell-k)!}\frac{(2m)!}{(2m+1-n+k)!}(-1)^\ell z^{2m+1-n+k+j+\ell-k}(2s)^{j-\ell}\notag\\
=&(n-1)!n! \lim_{z\rightarrow 0}\sum_{m=0}^n \sum_{j=0}^\infty \sum_{\ell=0}^j\frac{(-1)^\ell (2s)^{j-\ell}j!(j+\ell)!(2m)!}{m!\ell!(n-m)!(j-\ell)!}z^{2m+j+\ell+1-n}\notag\\
&\phantom{(n-1)!\lim_{z\rightarrow 0}}\times\sum_{k=0}^{n-1}\frac{1}{k!(n-k-1)!(j+\ell-k)!(2m+1-n+k)!}\notag\\
&=(-1)^{n+1} \lim_{z\rightarrow 0} \sum_{m=0}^n \sum_{j=0}^\infty \sum_{\ell=0}^j {j\choose \ell} {n\choose m} (-1)^\ell (2s)^{j-\ell}\frac{((n-1)-2m-\ell-j-1)!}{(-2m-\ell-j-1)!}z^{2m+j+\ell-(n-1)}
\end{align}
We can simplify the $z$-dependent term first:
\begin{align}
&\lim_{z\rightarrow 0}\frac{((n-1)-2m-\ell-j-1)!}{(-2m-\ell-j-1)!}z^{2m+j+\ell-(n-1)}\notag\\
=&\frac{(-1)!}{(-n)!}\delta(2m+j+\ell-(n-1))\notag\\
=&\lim_{x\rightarrow \pi} \frac{\pi}{\sin(x)}\frac{(n-1)!\sin(x n)}{\pi}\delta(2m+j+\ell-(n-1))\notag\\
=&(n-1)!(-1)^{n-1}\delta(2m+j+\ell-(n-1))
\end{align}
This simplifies the sum into the following form:

\begin{align}
&(-1)^{n+1} \lim_{z\rightarrow 0} \sum_{m=0}^n \sum_{j=0}^\infty \sum_{\ell=0}^j {j\choose \ell} {n\choose m} (-1)^\ell (2s)^{j-\ell}\frac{((n-1)-2m-\ell-j-1)!}{(-2m-\ell-j-1)!}z^{2m+j+\ell-(n-1)}\notag\\
=&(n-1)!\sum_{m=0}^n \sum_{j=0}^\infty \sum_{\ell=0}^j {j\choose \ell} {n\choose m}(-1)^\ell (2s)^{j-\ell}\delta(2m+j+\ell-(n-1))\notag\\
\equiv&\frac{1}{2^{n-1}}\sum_{j=0}^\infty \bigg\{((2s-1)^j)_\ell(2^n)_m \bigg\}\bigg|_{2m+j+\ell=n-1}
\end{align}
In the last line of the above, we have represented the integral as an infinite sum over two functions $(2s-1)^j$ and $2^n$, subjected to the constraint $2m+j+\ell=n-1$ on the indices of their binomial sum. The subscripts $\ell$ and $m$ represent which binomial indices correspond to which function. This concludes the derivation of a closed form for the residue of the integral at $z_1=0$. 

We will now calculate the contribution from the pole at $z_3=s-\sqrt{s^2-1}$:
\begin{align}
&\frac{1}{2^{n-1}}\textrm{Res}_{z\rightarrow z_3} \frac{(z^2+1)^n}{z^n \left(z-s-\sqrt{s^2-1}\right)\left(z-s+\sqrt{s^2-1}\right)}\notag\\
=&\frac{1}{2^{n-1}}\frac{
(s^2-2s\sqrt{s^2-1}+s^2-1+1)^n
}{
(s-\sqrt{s^2-1})^n (
-2\sqrt{s^2-1}
)
}\notag\\
=&\frac{1}{2^{n-1}}\frac{(2s^2-2s\sqrt{s^2-1})^n}{  
(s-\sqrt{s^2-1})^n(-2\sqrt{s^2-1}
}\notag\\
=&- \frac{s^n}{\sqrt{s^2-1}}
\end{align}
Taking into account the factor of $\eth(s)$ due to the pole at $z_3$ and taking the negative (physical) branch cut, we can now write down a final expression for the $n$th order Landau Chebyshev integral:
\begin{align}
\frac{1}{\pi}\int_{-1}^1 \frac{x^n}{s-x}\frac{dx}{\sqrt{1-x^2}}\equiv \mathcal{I}_n=\frac{s^n}{\sqrt{s^2-1}}\textrm{\dh}(s)-
\frac{1}{2^{n-1}}\sum_{j=0}^\infty \bigg\{((2s-1)^j)_\ell(2^n)_m \bigg\}\bigg|_{2m+j+\ell=n-1}
\end{align}

The above expression allows us to write the following form for the Landau kinetic equation in 2D, which is reproduced altogether below:

\begin{align}
\frac{1}{2}(1+\delta_{\ell 0}) \nu_\ell +\sum_{\ell'}\textrm{\th}_{\ell \ell'}(s)(1+\delta_{\ell' 0}) F_{\ell'}^s\nu_{\ell'} =-2\textrm{\th}_{\ell 0}(s)U
\end{align}
\begin{align}
\textrm{\th}_{\ell \ell'}(s)&=\frac{1}{2}\sum_{k=0}^{\floor{\frac{\ell'}{2}}}\sum_{j=0}^{\floor{\frac{\ell}{2}}} \sum_{m=0}^{k+j} {\ell' \choose 2k} {\ell \choose 2j} {k+j\choose m}(-1)^{m+1}\mathcal{I}_{\ell+\ell'-2m+1}
\end{align}
\begin{align}
\mathcal{I}_n=-\frac{s^n}{\sqrt{s^2-1}}\textrm{\dh}(s)+\frac{1}{2^{n-1}}\sum_{j=0}^\infty \bigg\{((2s-1)^j)_\ell(2^n)_m \bigg\}\bigg|_{2m+j+\ell=n-1}
\end{align}
\begin{align}
\textrm{\dh}(s)=1-\bigg\{
\delta(\Re(s))\Theta(|s|-1)+\delta(\Im(s))\Theta(1-|s|)
\bigg\}
\end{align}

Note that this differs significantly form the case of the 3D Fermi liquid, where the equation for the collective mode dispersion is in terms of the natural logarithms of $s$ (i.e., some infinite-order polynomial). This is seen more explicitly by recalling the general 3D Landau kinetic equation \cite{Baym}:

\begin{align}
\frac{\nu_\ell}{2\ell+1}+\sum_{\ell'} \Omega_{\ell \ell'}(s)F_{\ell'}^s \frac{\nu_{\ell'}}{2\ell'+1}=-\Omega_{\ell 0}(s)U
\end{align}
where $s=\omega/qv_F$ and $\Omega_{\ell \ell'}$ is given by

\begin{align}
\Omega_{\ell \ell'}(s)=\Omega_{\ell' \ell}(s)=\int_{-1}^1 \frac{d\mu}{2}P_\ell(x)\frac{x}{x-s}P_{\ell'}(x)
\end{align}

The form of $\Omega_{\ell \ell'}(s)$ determines the form of the sound dispersion. 
%
%
%
For some general $\ell$, the above can be written as can be written as
\begin{align}
\Omega_{\ell'\ell}(s)=\Omega_{\ell \ell'}(s)=\frac{\delta_{\ell \ell'}}{2\ell+1}-sP_{\ell'}(s)Q_\ell(s)
\end{align}
where $Q_{\ell}(s)$ is a Legendre function of the second kind, given by
\begin{align}
Q_\ell(s)=\frac{1}{s}\left(
-P_\ell(s)\Omega_{00}(s)+P_\ell(s)-sW_{\ell-1}(s)
\right)
\end{align}
where 
\begin{align}
W_{\ell-1}=\begin{cases}
& \sum_{k=1}^\ell\frac{1}{k}P_{k-1}(s)P_{\ell-k}(s),\quad \ell\ge 1\notag\\
&0,\quad \ell =0
\end{cases}
\end{align}
 From the above form, we can see that the dispersion of a collective mode in a three-dimensional Landau-Fermi liquid lacks a closed form valid for all values of $s$. The poles of the response function $\nu_0/U$ therefore depend strongly upon the form of $\Omega_{00}$, which is of the form of a natural logarithm. As such, the exact solution for the 3D zero sound dispersion $s$ lacks a closed form; we can only write down expressions in the limit $|s|<<1$ or $|s|>>1$.

\section{Conclusion}

In this article, we have extended the phase space analysis common in the description of the 3D Landau-Fermi liquid to the analogous 2D system. The underlying physical arguments for the existence of a Fermi liquid are the same, with the exception that the orthogonality condition on the mode expansion differs due to the inherent reduction of phase space. This has slight changes to the effective mass in the 2D Landau-Fermi liquid but has more interesting changes to the Pomeranchuk instability condition and the 2D Landau-Silin kinetic equation. In the case of the latter, the problem of finding the collective modes of the system at any arbitrary interaction strength (i.e., any finite set of Landau parameters $F_\ell$) reduces to finding the roots of some polynomial of finite order.

With the huge amount of interest in recent years on novel non-Fermi liquids in two dimensions, we hope our analysis sheds some light on how an unconventional Landau-Fermi liquid state can remain in some generic two-dimensional fermionic ensemble.

%
%
%

\section{Appendix A: Explicit derivation of Landau-Chebyshev integrals of the first, second, and third orders}

In this section, we will explicitly derive the Landau-Chebyshev integrals for $n=1,\,2$, and $3$. Recall that the general equation is given by

\begin{align}
&\frac{1}{\pi}\int_{-1}^1 \frac{x^n}{s-x}\frac{dx}{\sqrt{1-x^2}}
\notag\\
=&-\frac{s^n}{\sqrt{s^2-1}}\textrm{\dh}(s)+\frac{1}{2^{n-1}}\sum_{\substack{j=0 \\ \ell \le j}}^\infty (2s)^{j-\ell}(-1)^\ell{j\choose \ell}  \sum_{m=0}^n {n\choose m}\delta(2m+j+\ell-(n-1))
\end{align}
where we have written out the truncated binomial expansion explicitly. 

Taking $n=1$, the above simplifies nicely, as the second term converges to unity:

\begin{align}
&\frac{1}{\pi}\int_{-1}^1 \frac{x^1}{s-x}\frac{dx}{\sqrt{1-x^2}}
\notag\\
=&-\frac{s^1}{\sqrt{s^2-1}}\textrm{\dh}(s)+\frac{1}{2^{0}}\sum_{\substack{j=0 \\ \ell \le j}}^\infty (2s)^{j-\ell}(-1)^\ell{j\choose \ell}  \sum_{m=0}^1 {1\choose m}\delta(2m+j+\ell)\notag\\
=&\boxed{-\frac{s}{\sqrt{s^2-1}}\textrm{\dh}(s)+1}
\end{align}
This agrees with the previous result we derived in the text before we introduced the generalized the Landau-Chebyshev integral to higher orders.

We now calculate the above for $n=2$. 
\begin{align}
&\frac{1}{\pi}\int_{-1}^1 \frac{x^2}{s-x}\frac{dx}{\sqrt{1-x^2}}\notag\\
=&-\frac{s^2}{\sqrt{s^2-1}}\textrm{\dh}(s)+\frac{1}{2}\sum_{\substack{j=0 \\ \ell \le j}}^\infty (2s)^{j-\ell}(-1)^\ell{j\choose \ell}  \sum_{m=0}^2 {n\choose m}\delta(2m+j+\ell-1)\notag\\
=&
-\frac{s^2}{\sqrt{s^2-1}}\textrm{\dh}(s)+\frac{1}{2}
\left(
2s
\right)
\notag\\
=&\boxed{s\left(-\frac{s}{\sqrt{s^2-1}}\textrm{\dh}(s)+1\right)}
\end{align}
Unlike the case for $n=1$, it is interesting to note that the Landau-Chebyshev integral disappears as $s\rightarrow 0$ if $n=2$. 

Finally, we calculate the integral for $n=3$:

\begin{align}
&\frac{1}{\pi}\int_{-1}^1 \frac{x^3}{s-x}\frac{dx}{\sqrt{1-x^2}}\notag\\
&=-\frac{s^3}{\sqrt{s^2-1}}\textrm{\dh}(s)+\frac{1}{2^{2}}\sum_{\substack{j=0 \\ \ell \le j}}^\infty (2s)^{j-\ell}(-1)^\ell{j\choose \ell}  \sum_{m=0}^3 {3\choose m}\delta(2m+j+\ell-2\notag\\
=&
\boxed{-\frac{s^3}{\sqrt{s^2-1}}\textrm{\dh}(s)+\frac{1}{2}(2s^2+1)}
\end{align}
Without loss of generality, we can see that higher-orders of the Landau-Chebyshev integral are given by the first term dependent on \dh$(s)$ plus some $(n-1)$th order polynomial of $s$.

\newpage 
\bibliographystyle{iopart-num}
\bibliography{main}{}

\end{document}